# Gate-Tunable Spin-to-Charge Conversion in Topological Insulator-Magnetic Insulator Heterostructures at Room Temperature

Wenxuan Sun,[1] Yequan Chen,[1] Ruijie Xu,[1] Wenzhuo Zhuang,[1] Di Wang,[2] Long Liu,[2] Anke Song,[1] Guozhong Xing,[2,*] Yongbing Xu,[1] Rong Zhang,[1,3] Cui-Zu Chang,[4] Xuefeng Wang[1,5,*]

[1]Jiangsu Provincial Key Laboratory of Advanced Photonic and Electronic Materials, State Key Laboratory of Spintronics Devices and Technologies, School of Electronic Science and Engineering, Collaborative Innovation Center of Advanced Microstructures, Nanjing University, Nanjing 210093, China

[2]Institute of Microelectronics, Chinese Academy of Sciences, Beijing 100029; University of Chinese Academy of Sciences, Beijing, 100049, China

[3]Department of Physics, Xiamen University, Xiamen 361005, China

[4]Department of Physics, The Pennsylvania State University, University Park, PA 16802, USA

[5]Atom Manufacturing Institute, Nanjing 211806, China

*Authors to whom correspondence should be addressed.
E-mail: gzxing@ime.ac.cn (G. Xing); xfwang@nju.edu.cn (X. Wang)

## Abstract

Over the past decade, topological insulators have received enormous attention for their potential in energy-efficient spin-to-charge conversion, enabled by strong spin-orbit coupling and spin-momentum locked surface states. Despite extensive research, the spin-to-charge conversion efficiency, usually characterized by the spin Hall angle ($\theta_{SH}$), remains low at room temperature. In this work, we employed pulsed laser deposition to synthesize high-quality ternary topological insulators $(Bi_{0.1}Sb_{0.9})_2Te_3$ thin films on magnetic insulator $Y_3Fe_5O_{12}$. We find that the value of $\theta_{SH}$ reaches ~0.76 at room temperature and increases to ~0.9 as the Fermi level is tuned to cross topological surface states via electrical gating. Our findings provide an innovative approach to tailoring the spin-to-charge conversion in topological insulators and pave the way for their applications in energy-efficient spintronic devices.

## 1. Introduction

The spin-orbit torque (SOT), which is generated by the spin accumulation in nonmagnetic layers via spin-orbit coupling (SOC), facilitates efficient manipulation and switching of the magnetization in the adjacent ferromagnetic layers,[1-4] making it a promising mechanism for the development of memory and logic devices.[5-8] A higher conversion efficiency from spin current to charge current [i.e., spin-to-charge conversion (SCC)] is crucial for the potential practical applications in SOT devices and is typically characterized by the spin Hall angle ($\theta_{SH}$). Over the past decade, there has been a notable shift in focus from heavy metals[9-12] to alternative spin-to-charge converters aimed at enhancing $\theta_{SH}$. These include topological semimetals,[13-15] transition metal dichalcogenides,[16-18] two-dimensional (2D) materials,[19-20] antiferromagnets[21-22] and 2D electron gases at Rashba interfaces.[23-24] In particular, topological insulators (TIs) have demonstrated relatively large $\theta_{SH}$ due to the spin-momentum locking of their topological surface states.[25-26] This property is often assessed through spin-torque ferromagnetic resonance,[27-30] spin-pumping ferromagnetic resonance (SP-FMR),[31-45] and second harmonic measurements.[46-47] However, defects in TIs often lead to their bulk conductivity, resulting in unintended participation in the SCC process and a general low $\theta_{SH}$ at room temperature.[48-50] Thus, it is urgent to tailor the bulk and surface states to enhance the $\theta_{SH}$.

Electric-field control is a vital knob to manipulate the performance of electronic devices[51-55] by tuning the Fermi level of TIs.[56-59] It has been demonstrated that the Fermi-level-dependent $\theta_{SH}$ could be adjusted by varying the thickness[50] and the composition[29, 35, 37] of the TIs. Recent studies have also shown the effective manipulation of SOT strength using an electric field at low temperatures, as measured by second harmonic[47, 60] and hysteresis loop shift methods.[61] A constant $\theta_{SH}$ was observed in SCC experiments when the Fermi level was tuned within the band gap under the accessible gate voltage at 50 K.[35] Furthermore, the effective control of the SP-FMR process was demonstrated by investigating the dissipation of angular momentum at 10 K, where the Fermi level was tuned across the band gap and touched the bulk conduction band and valence band.[62] However, achieving electric tuning of $\theta_{SH}$ in TI-based heterostructures at higher temperatures remains elusive.

In this work, we report an enhanced $\theta_{SH}$ in ternary $(Bi_{0.1}Sb_{0.9})_2Te_3$ (BST) thin films grown on yttrium iron garnet (YIG, $Y_3Fe_5O_{12}$) by the pulsed laser deposition (PLD) method. The robust $\theta_{SH}$ reaches as large as ~0.76 at room temperature, as determined by SP-FMR measurements. Moreover, the value of $\theta_{SH}$ further increases to ~0.9 upon electrical gating, which is attributed to the contribution of topological surface states as the Fermi level is tuned. This work provides an in-depth insight into the surface state contribution to $\theta_{SH}$ and facilitates the applications in the dissipationless spintronic devices.

## 2. Results and Discussion

The BST thin films were epitaxially grown on liquid-phase-epitaxial magnetic insulator YIG using the PLD method (Experimental Section). The BST films are structured in repeated quintuple layers of atoms (Te-*M*-Te-*M*-Te, where *M* = Bi or Sb), stacking along the *c*-axis (**Figure 1**a).[63] Figure 1b demonstrates a scanning transmission electron microscopy (STEM) image of the BST (15 nm)/YIG heterostructure, which reveals atomically ordered layers (Figure S1, Supporting Information). An amorphous YIG layer approximately 3 nm thick is observed at the interface, likely resulting from the substrate polishing process. Moreover, the energy-dispersive X-ray spectroscopy (EDX) elemental mapping (Figure 1c) presents a sharp interface without the obvious elemental interdiffusion. The stoichiometry of Bi:Sb determined from the EDX elemental mapping is found to be about 1:9, in agreement with the designed composition of the target material. Figure 1d shows a representative $\theta$-$2\theta$ X-ray diffraction (XRD) pattern of the BST films, where the preferential direction along the $c$-axis is distinct, verifying the high-quality crystalline structure. Additional characterization from Raman spectroscopy and X-ray photoemission spectroscopy (XPS) consistently reveals the well-formed structure of BST films (Figure S2, Supporting Information). In comparison, the intensities of the Raman spectra and XRD patterns for BST/sapphire heterostructures are stronger with high signal-to-noise ratio, indicating a better crystalline structure (Figure S2a,b, Supporting Information). Oxidation states are observed in all elements in the XPS spectra, which is induced from the contamination during the transferring process (Figure S2c, Supporting Information). By estimating the area under Bi 4f and Sb 3d levels, the atomic concentration ratio of Bi:Sb is 0.08:0.92, very close to the result of EDX elemental mapping.

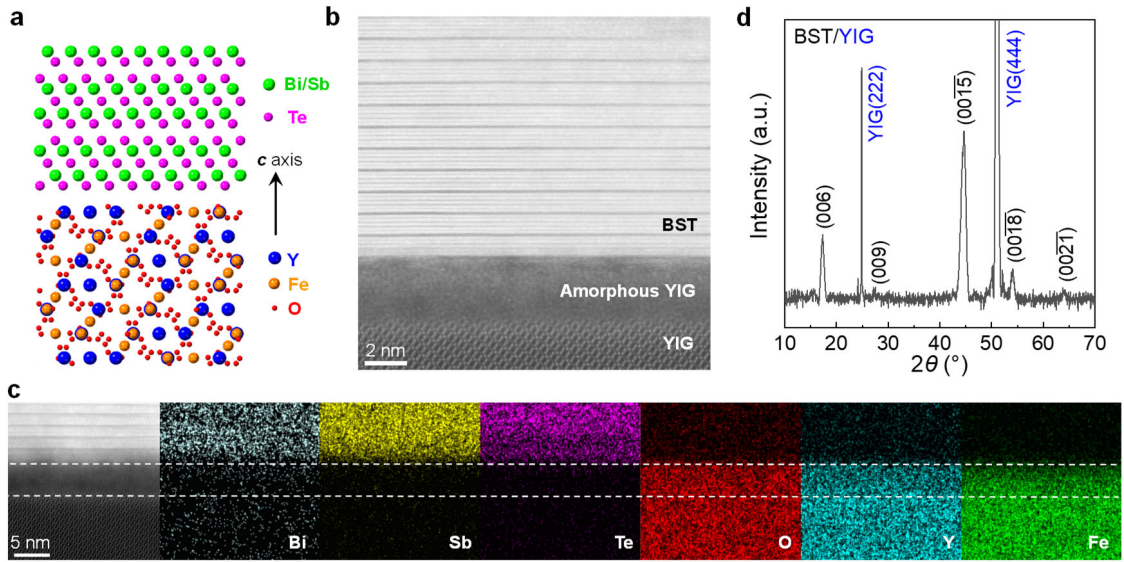

**Figure 1.** Structural characterization of the 15-nm-thick BST films grown on YIG. a) Schematic of the BST/YIG crystalline structure. b) Cross-sectional STEM image of the BST/YIG heterostructure. c) The corresponding EDX elemental maps of Bi, Sb, Te, O, Y, and Fe, respectively. d) The $\theta$-$2\theta$ XRD pattern of the BST/YIG heterostructure.

Next, we perform SP-FMR measurements to investigate the SCC in BST/YIG heterostructures at room temperature. **Figure 2**a exhibits the schematic diagram of SP-FMR measurements. As the samples are attached to the coplanar waveguide (CPW), a microwave field of gigahertz (GHz), generated by the radio frequency (RF) current through the signal line of CPW, induces a precession of the magnetization in YIG layers (Figure S3, Supporting Information). At the resonance condition, spin current $J_s$ is generated in YIG and injected into the BST layers via FMR. The strong SOC of BST subsequently converts the spin current into a charge current $J_c$, leading to the critical spin pumping signals. During the measurements, an external magnetic field ($H$) is applied along the $x$-axis, and the total dc-voltage ($V_{total}$) is measured across the BST layers along the $y$-axis.

Figure 2b shows field-dependent $V_{total}$ measured at various frequencies ($f$ = 4-9 GHz) for the 6-nm-thick BST films using 15 mW microwave power. Pronounced peaks are observed at the corresponding resonance fields, and there is a change in sign with the reversed direction of $H$, as expected in SP-FMR measurements. A representative $V_{total}$ measured at $f$ = 5 GHz (Figure 2c) consists of a symmetric and an antisymmetric Lorentzian function component (marked with green and blue lines, respectively). The experimental data are fitted by the following equation to extract the symmetric and antisymmetric components:[64]

$$V_{\text{total}} = V_S \frac{\Delta H^2}{\Delta H^2 + (H-H_r)^2} + V_A \frac{\Delta H(H-H_r)}{\Delta H^2 + (H-H_r)^2} \tag{1}$$

where $\Delta H$ is the linewidth, and $H_r$ is the resonance field. The coefficient of the symmetric part ($V_S$) is the voltage signal arising from the SP-FMR and the Seebeck effect. In contrast, the coefficient of the antisymmetric part ($V_A$) stems from the contribution of the anisotropic magnetoresistance or the anomalous Hall effect. We note that the SP-FMR voltage ($V_{sp}$) reverses its sign at positive or negative fields while the Seebeck voltage does not change. Consequently, the $V_{sp}$ can be separated from the Seebeck voltage by using $V_{sp} = \frac{V_S(+H_r) - V_S(-H_r)}{2}$.

The 3D charge current density $J_C^{3D}$ is then obtained by: $J_C^{3D} = \frac{V_{sp}}{Rwt_{BST}}$, where $R$, $w$ and $t_{BST}$ are the resistance (measured by the two-terminal method), the width of samples and the thickness of BST films, respectively.

Figure 2d presents the resonant-field-dependent frequency, which is well fitted by the Kittel formula:[65]

$$f = \frac{\gamma}{2\pi}\sqrt{H_r(H_r + 4\pi M_{eff})} \tag{2}$$

where $\gamma$ is the gyromagnetic ratio, and $M_{eff}$ is the effective saturation magnetization. Figure 2e shows the frequency-dependent linewidth of the BST/YIG heterostructure (red) and the bare YIG substrate (black). The damping factor $\alpha$ is obtained by fitting the data according to the relation $\Delta H = \Delta_0 + \frac{2\pi}{\gamma}\alpha f$, where $\Delta_0$ is the inhomogeneous line broadening factor. An enhancement of $\alpha$ is observed in the BST/YIG heterostructure ($2.4 \times 10^{-3}$) compared to the bare YIG ($0.8 \times 10^{-3}$ obtained from FMR absorption), revealing the generation and absorption of spin current. Furthermore, the corresponding effective spin mixing conductance $g_{\uparrow\downarrow}$ (i.e., the efficiency of interfacial spin transport) is calculated by the following equation:[66]

$$g_{\uparrow\downarrow} = \frac{4\pi M_s t_{YIG}}{g\mu_B}(\alpha_{BST/YIG} - \alpha_{YIG}) \tag{3}$$

where $g$, $\mu_B$, $M_s$, and $t_{YIG}$ are Landé g-factor, Bohr magneton, saturation magnetization, and coherence length of the YIG layer (300 nm)[67], respectively. The $g_{\uparrow\downarrow}$ is ~$6.9 \times 10^{18}$ m$^{-2}$ in the BST/YIG heterostructure, which may be slightly lowered due to the amorphous YIG layer[68]. The spin current density injected into the BST/YIG interface is provided by:[66]

$$J_S = \frac{g_{\uparrow\downarrow}\gamma^2 h_{rf}^2 \hbar}{8\pi\alpha^2} \frac{(4\pi M_s\gamma + \sqrt{(4\pi M_s\gamma)^2 + 4\omega^2})}{(4\pi M_s\gamma)^2 + 4\omega^2} \frac{2e}{\hbar} \tag{4}$$

where $h_{rf}$ is the induced microwave RF field, calculated from Ampere's law (10 mOe), $\omega = 2\pi f$ represents the excitation frequency, and $\hbar$ is the reduced Planck constant. To further figure out the origin of the SCC in BST/YIG heterostructures, a series of samples with varying thicknesses of the BST layer are prepared, on which SP-FMR measurements are implemented following the steps mentioned above. The measured BST-thickness-dependent $V_{sp}/R$ is exhibited in Figure 2f. The $V_{sp}/R$ increases with increasing thickness and tends to become saturated, which is a typical feature of the spin diffusion model expressed by: [17]

$$\frac{V_{sp}}{R} = w\theta_{SH}\lambda_{SD}\tanh\left(\frac{t_{BST}}{2\lambda_{SD}}\right)J_S \quad (5)$$

where $\theta_{SH}$ and $\lambda_{SD}$ represent the SCC efficiency (i.e., spin Hall angle) and the spin diffusion length, respectively. We obtain $\theta_{SH} = 0.76$ with $\lambda_{SD} = 6.25$ nm by fitting the experimental data to the formula above. The spin diffusion length is very close to that reported in $Bi_2Se_3$/Py heterostructures,[69] while the $\theta_{SH}$ demonstrates an enhancement of two orders of magnitude. The measured $V_{sp}$ values of the BST/YIG and Pt/YIG heterostructures possess the same polarity, indicating both positive signs of $\theta_{SH}$ in BST and Pt (Figure S4, Supporting Information). Several recent works have reported the highly efficient SCC in TI-based heterostructures via the inverse Edelstein effect (IEE), which is characterized by the inverse Edelstein length:[38]

$$\lambda_{IEE} = \frac{J_C^{2D}}{J_S} = \theta_{SH}\lambda_{SD} \quad (6)$$

Using $\lambda_{SD} = 6.25$ nm, we obtain $\theta_{SH}$ of 0.34, 0.32, 0.1 and 0.13 for α-Sn,[38] HgTe,[39] $Sb_2Te_3$[40] and $BiSbTe_{1.5}Se_{1.5}$,[34] respectively. The $\theta_{SH}$ (0.76) in our work is clearly higher than these values, indicating the large SCC efficiency in BST films. The equivalent $\lambda_{IEE}$ of BST/YIG is also calculated according to Equation (6), which is of the same order of magnitude as the reported values (Table S1, Supporting Information).

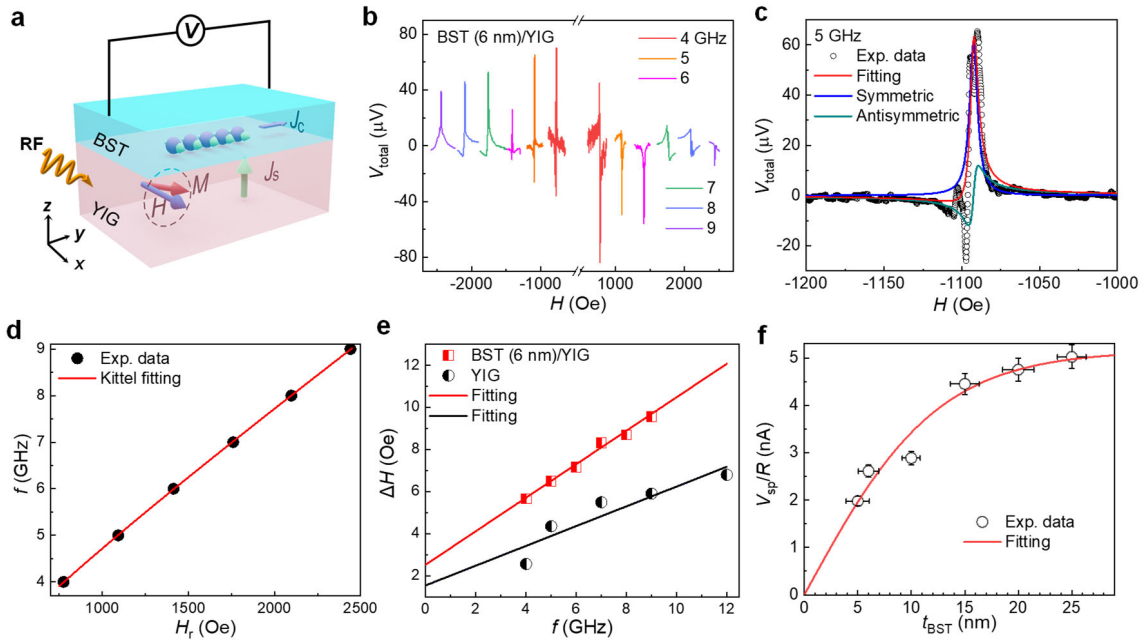

**Figure 2.** SP-FMR results of BST/YIG heterostructures. a) Schematic diagram of the SP-FMR measurements. b) The magnetic field-dependent total dc-voltage ($V_{total}$) measured on the 6-nm-thick BST films at different RF frequencies. c) A representative magnetic field-dependent $V_{total}$ at 5 GHz fitted by Lorentzian function. d) The resonant-field-dependent frequency by the Kittel fitting. e) The frequency-dependent linewidth of the BST/YIG heterostructure and the YIG substrate, respectively. f) The BST-thickness-dependent $V_{sp}/R$ fitted by the spin diffusion model, which determines the value of $\theta_{SH}$.

Recent research on TIs has demonstrated that the carrier density of the topological surface states can significantly affect the value of $\theta_{SH}$, which highly relies on the position of the Fermi level.[29, 61, 70] This inspires us to electrically tailor the Fermi level of BST to obtain a higher $\theta_{SH}$. A 15-nm-thick $Al_2O_3$ layer is deposited on the 6-nm-thick BST films as the top-gate dielectric layer by atomic layer deposition (**Figure 3**a). Figure 3b shows the field-dependent $V_{total}$ measured upon gate voltages ($V_g$ = -2-3 V) with a fixed microwave frequency of 6 GHz. Figure 3c exhibits the corresponding field-dependent symmetric part $V_{symmetric}$ obtained from the $V_{total}$ following the steps as before. Figure 3d provides the gate-voltage-dependent $V_{sp}$. When negative $V_g$ is applied, a notable reduction of $V_{sp}$ is observed. As $V_g$ sweeps from 0 V to 3 V, $V_{sp}$ reaches a maximum of ~48.1 µV at $V_g$ = 1 V. Afterwards, $V_{sp}$ decreases with increasing $V_g$. The extracted $\Delta H$ and $H_r$, related to the generated $J_S$, remain almost unchanged, indicating the negligible change in $J_S$ upon the electrical gating.

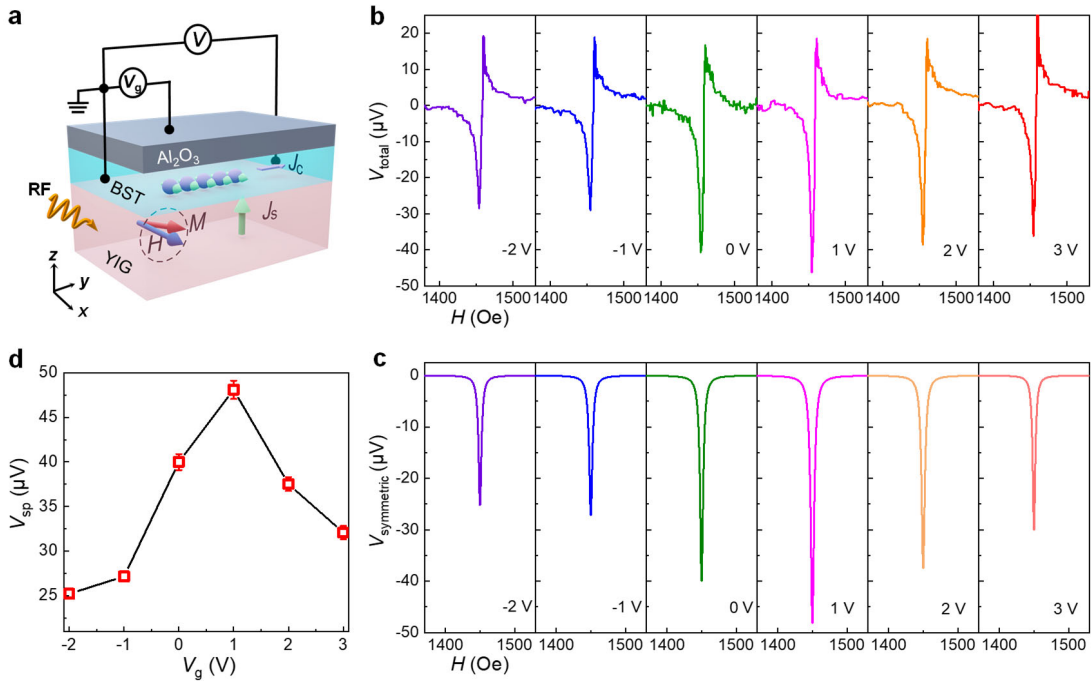

**Figure 3.** Gate tunable SCC in the BST (6 nm)/YIG heterostructure. a) Schematic diagram of the gate tunable SCC setup. b) The magnetic-field-dependent $V_{total}$ under different gate voltages. c) The magnetic-field-dependent symmetric component ($V_{symmetric}$) of the $V_{total}$ under different gate voltages. d) The gate-voltage-dependent $V_{sp}$, which shows a maximum at $V_g$ = 1 V.

To gain more insights into the mechanism of the electric-field control of the SCC process, Hall measurements are conducted upon different $V_g$, in which one could achieve the evolution of the Fermi level during the electrical gating. The extracted gate-voltage-dependent carrier density monotonously decreases as $V_g$ sweeps from -2 to 3 V (**Figure 4**a). One should note that the electrical transport reveals the uniform behavior of the conductive layer. At the same time, the SCC is more sensitive to the interface condition.[71] The top-gating induces the Fermi level shifting of BST at the interface with the $Al_2O_3$ layer. The band structure along the *c*-axis can be viewed as a lever arm, resulting in a slight shift of the Fermi level at the bottom surface upon gating.[62] This process enables the precise control of the Fermi level at the interface between BST and YIG, leading to the electrically modulated $\theta_{SH}$. Note that the top-gating significantly influences the value of $\theta_{SH}$, which shows the maximum value of 0.9 at $V_g = 1$ V (Figure 4b).

Next, we explore the primary physical origin of the SCC in BST/YIG heterostructures. The temperature-dependent resistivity of the BST films reveals a metallic behavior with the resistivity decreasing at lower temperatures (see Figure S5a, Supporting Information). Additionally, the density of the majority carrier holes (~$10^{20}$ cm$^{-3}$) is obtained from the Hall measurement at room temperature (see Figure S5b, Supporting Information). These are the typical characteristics of TIs with elevated tellurium content,[72-73] consistent with our EDX results (the stoichiometric ratio of Te is 63%). Normally, there are two main intrinsic defects in the as-grown Bi$_2$Te$_3$ crystals, i.e., Bi-on-Te antisites and Te vacancies; the former provides holes for the hole transport and the latter facilitates the electron transport.[74-75] The slightly elevated Te content suppresses the formation of Te vacancies, driving the bulk states towards p-type conductivity. Meanwhile, increasing the Sb content in the p-type (Bi,Sb)$_2$Te$_3$ promotes the formation of Sb-on-Te antisites, which act as electron acceptors.[73, 75] These intrinsic defects induced during growth act as hole doping, leading to the p-type bulk conductivity. Consequently, the Fermi level is located in the valence band, leading to bulk states dominating the transport properties. As a result, a significant portion of the generated spin current flows through the bulk states of BST layers during the SP-FMR process, converting to charge current via the inverse spin Hall effect (ISHE). This observation agrees with the well-fitted spin diffusion model (Figure 2f) and the $\sin^2\varphi$ dependence of the $V_{sp}$ on precession cone angle $\varphi$ (Figure S6, Supporting Information). The topological surface states are expected to produce the higher $\theta_{SH}$ than bulk states due to the spin-momentum locking. Accordingly, an appropriate stoichiometry of Bi:Sb (1:9) is reported to possess the most dominant topological surface states.[76-77] The weak antilocalization at low temperatures indicates the existence of the strong SOC in the topological surface states (Figure S5c,d, Supporting Information). The angle-resolved photoelectron spectroscopy (ARPES) result further confirms the persistence of the bulk energy gap and topological surface states at room temperature (Figure S7, Supporting Information). Therefore, the IEE at the interface also contributes to the SCC owing to the topological surface states, giving rise to the large $\theta_{SH}$.

The gate-voltage-dependent $\theta_{SH}$ and carrier density in the BST/YIG heterostructures are directly related to the evolution of the Fermi level (Figure 4a). At $V_g = 0$ V, the Fermi level is located in the valance band at both top and bottom surfaces so that the primary effect of the SCC is the ISHE in the conductive bulk states. When negative $V_g$ is applied, the Fermi level moves deeper into the valence band, increasing bulk carrier density. This bulk dominance reduces $\theta_{SH}$ due to the enhanced scattering probabilities.[62] When applying positive $V_g$, the Fermi level moves towards the band gap and reaches the valence band maximum at $V_g = 1$ V. The IEE in the topological surface states dominates the SCC, and $\theta_{SH}$ reaches the maximum. Subsequently, as $V_g$ continues to increase, the density of states (DOS) in the surface states significantly decreases with the Fermi level approaching the Dirac point, resulting in a reduction of the number of electrons involved in the SCC process.[62] Moreover, the scattering induced by inhomogeneities may reduce the generated spin polarization, as reported in the previous work.[29] In the surface states near the Dirac point, electrons with the lower energy are more susceptible to scattering from impurities and defects, leading to increased dissipation of the spin current and a reduction of the $\theta_{SH}$. Since the thickness of the BST film (6 nm) is very close to the spin diffusion length, the contribution from the top surface states is not considered. The majority carrier type does not change during the electrical gating, indicating that the Fermi level remains below the Dirac point. Although the evolution of the Fermi level does not cover the whole band diagram, our results demonstrate that electrical gating is a powerful approach to manipulate the bulk and surface states that contribute to the $\theta_{SH}$.

We summarize the $\theta_{SH}$ obtained via SP-FMR experiments in various heterostructures (Figure 4c), in which nonmagnetic materials are explored as spin-to-charge convertors.[9, 16-18, 41, 43] The SCC process in these heterostructures is attributed to the bulk ISHE. It is seen that the BST/YIG heterostructures in our work possess the highest $\theta_{SH}$ at room temperature, which highlights the crucial contribution of the topological surface states. Jamali et al. reported that the surface-states-induced SOC can enhance the spin current pumped into the bulk states, resulting in a giant $\theta_{SH}$ of 0.43 in $Bi_2Se_3$.[41] In this work, we deposit BST films on ferromagnetic substrates, ensuring a clear and sharp interface as evidenced by the STEM images (Figure 1b). Moreover, the 3-nm-thick amorphous YIG layer prevents the topological surface states from contacting directly with the ferromagnetic layers, thereby maximizing the strong SOC induced by the topological surface states and thus achieving a high $\theta_{SH}$. Taking advantage of the gate-tunable band diagram of the BST layers, $\theta_{SH}$ exhibits an extensive range of regulation under electrical gating, which could find potential applications in gate-controlled spintronic devices.

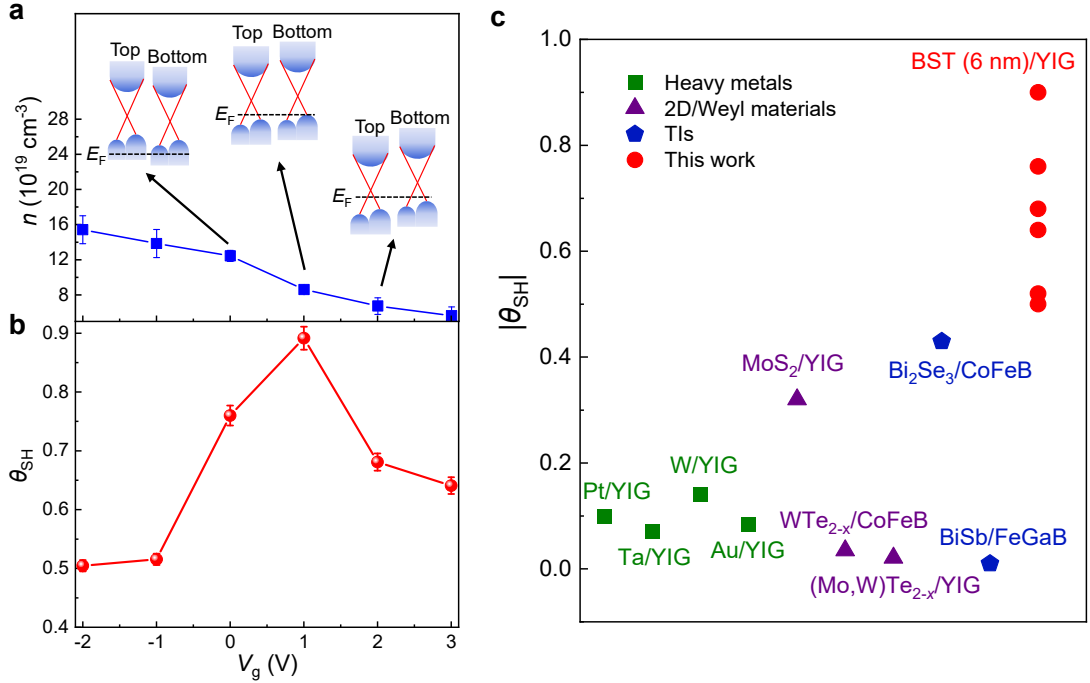

**Figure 4.** Enhanced $\theta_{SH}$ in BST/YIG heterostructures at room temperature. a,b) Gate-voltage-dependence of the carrier density and $\theta_{SH}$, respectively. The inset of (a) shows the Fermi level ($E_F$) evolution of BST under different gate voltages. c) Summary of the absolute values of $\theta_{SH}$ determined by SP-FMR for BST (this work) and other nonmagnetic materials, such as heavy metals,[9] 2D/Weyl materials,[16-18] and TIs,[41, 43] which highlights a large and gate-tunable $\theta_{SH}$ of BST in this work.

## 3. Conclusion

In summary, we have successfully fabricated high-quality epitaxial BST films on YIG substrates using the PLD method. SP-FMR measurements are employed to investigate the SCC in BST/YIG heterostructures. A substantial $\theta_{SH}$ of ~0.76 is achieved at room temperature, which is attributed to the ISHE in the bulk states associated with the IEE in the topological surface states. By applying a top-gate, we could tune the Fermi level from the valance band into the band gap, further enhancing the value of $\theta_{SH}$ to 0.9. The strategy of the enhanced $\theta_{SH}$ in BST/YIG heterostructures opens up a new avenue for tailoring the SCC in TIs through electrical gating for energy-efficient spintronic devices.

## 4. Experimental Section

*Growth of BST-Based Heterostructures:* The high-quality BST films with various thicknesses were prepared using the PLD method on the liquid-phase-epitaxial YIG and (0001)-oriented sapphire substrates, respectively. The target used for deposition was synthesized by mixing highly pure $Bi_2Te_3$ (99.99%) and $Sb_2Te_3$ (99.99%) powders with a stoichiometric ratio of 1:9. The mixture was then sealed in an evacuated quartz tube and heated at 700 °C for a week. The thoroughly reacted powder was pressed into a hard target using a Dry Powder Press Machine, which was further sintered at 650 °C for 3 days. Before deposition, the YIG substrates were cleaned and loaded into the PLD vacuum chamber (with a base pressure of ~1 × 10$^{-7}$ mbar) parallel to the BST target at a distance of ~5 cm. During the growth process, the target was ablated using a 248 nm KrF excimer laser with an average fluence of ~1.5 J cm$^{-2}$ and a repetition rate of 2 Hz when the substrates were kept at 200 °C. The $Al_2O_3$ acting as dielectric layers were deposited onto the BST/YIG heterostructures via the atomic layer deposition method at 150 °C.

*Structural Characterization:* The crystalline structure was characterized by the $\theta-2\theta$ XRD using Cu $K_\alpha$ radiation (Rigaku D/MAX-Ultima III). STEM and EDX mapping (FEI Titan Themis 200 TEM) were employed to investigate the microstructure and elemental distribution of the heterostructures. The microscope was operated at 200 kV accelerating voltage. The crystallography was further determined by a micro-Raman spectrometer (NT-MDT nanofinder-30) and an x-ray photoelectron spectroscopy from ULVAC-PHI 5000 Versa Probe using Al $K_\alpha$ source.

*SP-FMR and Hall Measurements with Applied Top-Gate:* For SP-FMR measurements, the heterostructures were patterned into stripes of 3.8 mm length and 400 μm width using the hard mask, which were fixed onto a coplanar waveguide (CPW) with adhesive tape (the substrates were located close to the CPW surface). The CPW comprised a 1 mm-wide signal line between two ground lines with a gap of 0.1 mm wide, connected to a broadband microwave generator (1-20 GHz). The microwave frequency was fixed during the measurements, and the external magnetic field was swept. An electromagnet was employed to apply the external magnetic field, and the voltage signals were detected using a Keithley 2182A nanovoltmeter. Hall measurements were implemented in the standard Hall-bar geometry. A Cryogen Free Measurement System (CFMS, Cryogenic) was employed with the perpendicular magnetic field up to 3.5 T. During the electrical gating process, the gate voltages were applied using a Keithley 2400 sourcemeter. The SP-FMR and Hall measurements were performed at fixed $V_g$. Between acquisitions, $V_g$ was changed monotonously from -2 to 3 V.

**Acknowledgements**

We thank Prof. Lichuan Jin for providing YIG substrates. This work was supported by the National Key Research and Development Program of China (Grant Nos. 2022YFA1402404 and 2021YFB3601300), the National Natural Science Foundation of China (Grant Nos. T2394473, 62274085, 11874203, 61822403 and 62374180), the Fundamental Research Funds for the Central Universities (Grant No. 021014380225), and the Strategic Priority Research Program of the Chinese Academy of Sciences (Grant No. XDB44010000). C.-Z.C. acknowledges the support from the Gordon and Betty Moore Foundation's EPiQS Initiative (Grant No. GBMF9063) and the NSF (Grant No. DMR-2241327).

# Supporting Information

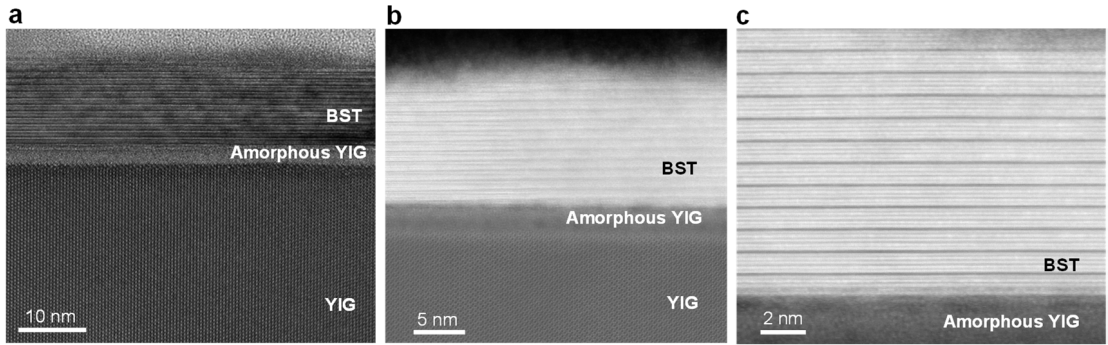

**Figure S1.** Cross-sectional STEM images of the BST (15 nm)/YIG heterostructure with different magnifications, revealing the high-quality crystalline structure and the homogeneity on a large scale. It has been demonstrated that PLD is a suitable method for high-quality growth of various materials.[1] Our work shows that PLD is also an outstanding technique for fabricating large-scale topological insulators. With the advantages of high stoichiometric fidelity, rapid growth pace and precise thickness control, PLD method shows great potential in designing and fabricating various heterostructures with large scale, high crystallinity, and good uniformity, especially in the development of TIs-based devices.

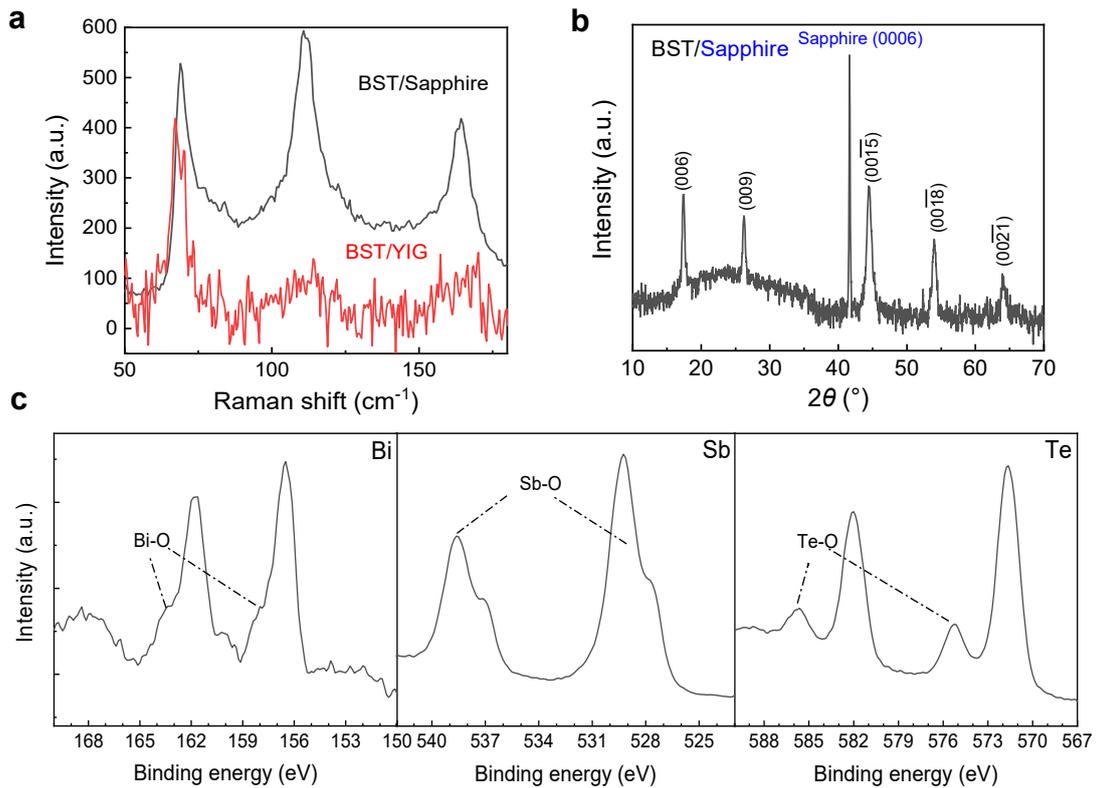

**Figure S2.** Additional structural characterization of the 15-nm-thick BST films grown on different substrates. a) Raman spectra of the 15-nm-thick BST films grown on the YIG and sapphire substrates. Pronounced peaks are observed at 68.94, 110.75 and 163.66 cm$^{-1}$, corresponding to the $A_{1g}^1$, $E_g^2$, $A_{1g}^2$ modes, respectively. b) XRD pattern of the BST (15 nm)/sapphire heterostructure, exhibiting diffraction peaks of (003$n$) family of planes. c) High-resolution XPS spectra of Bi, Sb and Te, respectively.

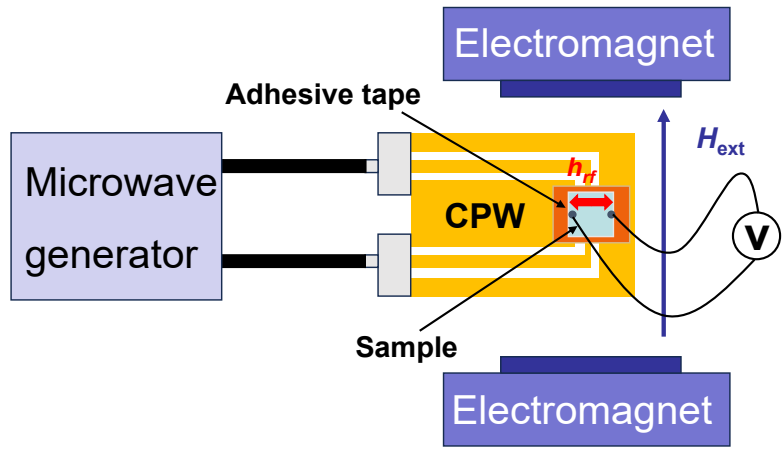

**Figure S3.** Detailed schematic diagram of the experimental setup for the SP-FMR measurements.

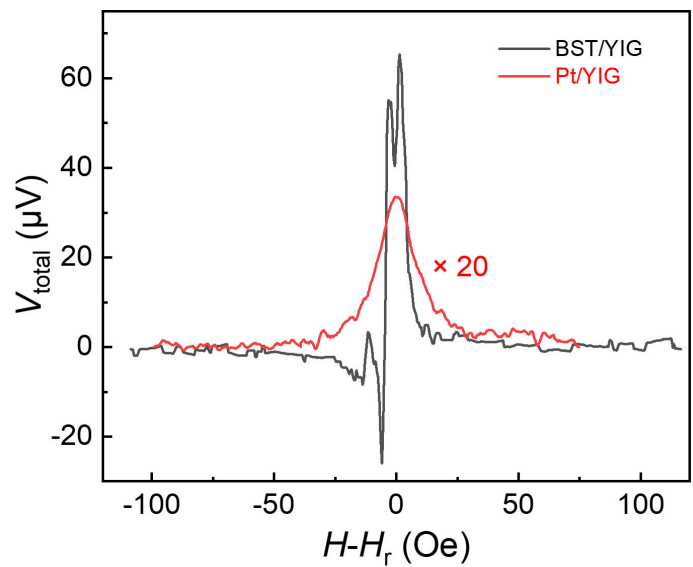

**Figure S4.** The spin pumping signals of the BST (6 nm) and Pt (1 nm) films grown on YIG substrates. The spin pumping signals measured in BST and Pt have the same polarity at the positive field, indicating that the sign of $\theta_{SH}$ in BST is the same as that in Pt (i.e., positive).

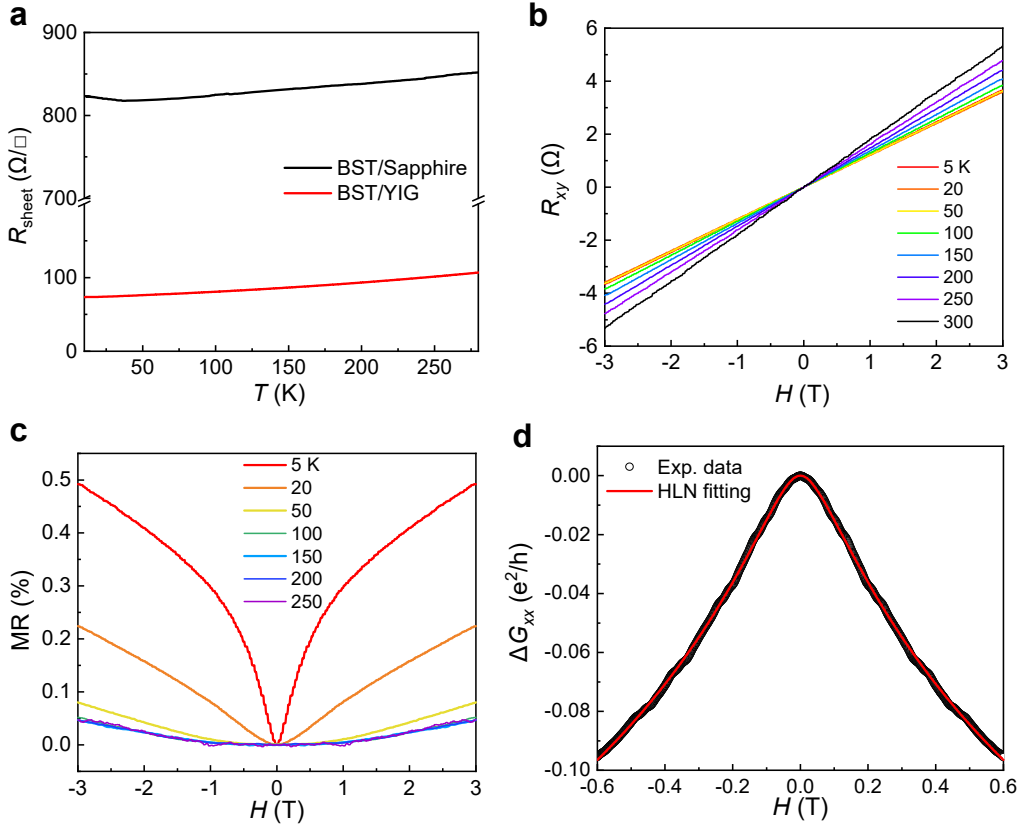

**Figure S5.** Transport properties of the BST (10 nm)/YIG heterostructure. Since the substrates are insulating, the electrical signals totally come from the BST films. a) The temperature-dependent sheet resistance of the 10-nm-thick BST films grown on the YIG and sapphire substrates. Both heterostructures show a metallic behavior that the resistance decreases at low temperature, indicating that intrinsic defects are nonnegligible. Meanwhile, the BST film grown on the YIG substrate is more conductive compared to that on the sapphire. b) The Hall curves of the BST/YIG heterostructure at various temperatures. The field-dependent Hall resistance of the BST/YIG heterostructure shows p-type behavior at all temperatures from 5 K to 300 K, which confirms the conductive bulk states of the BST films on YIG. Consequently, most of the generated spin current flows into the bulk of BST, leading to the inverse spin Hall effect governing the SCC. c) The magnetoresistance curves of the BST/YIG heterostructure at various temperatures. A sharp cusp in the magnetoresistance curve at low fields is observed at 5 K, which is the characteristic feature of the weak antilocalization effect (WAL).[2] Owing to the spin momentum-locking Dirac cone, the surface state of BST can acquire a π Berry phase, which induces a quantum enhancement to the classical electronic conductivity and leads to the WAL.[3] d) HLN fitting of the weak antilocalization effect at 5 K. The magnetoconductivity on the surface states of a topological insulator can be expressed by the Hikami-Larkin-Nagaoka (HLN) formula:[4]

$$\Delta G_{xx}(B) = \alpha e^2/(2\pi^2 \hbar)[\ln\left(\frac{B_\varphi}{B}\right) - \Psi(\frac{1}{2} + \frac{B_\varphi}{B})],$$

where $B_\varphi = \hbar/(4eL_\varphi^2)$ is a characteristic field defined by the dephasing length $L_\varphi$, $\psi$ is the digamma function, and $\alpha$ is a constant. $L_\varphi$ = 161 nm with $\alpha$ = 0.29 is obtained by fitting the data (Figure S5d). The value of α = 0.29 indicates the presence of topological surface states in the BST thin films and contribution of the bulk states to the transport properties.

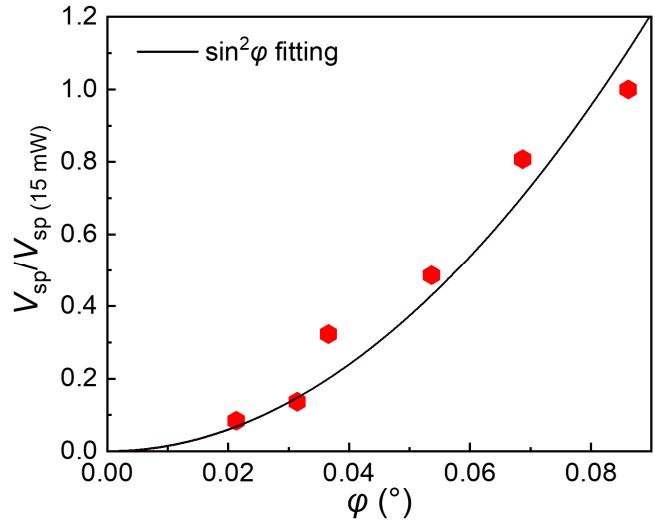

**Figure S6.** The precession-cone-angle-dependent $V_{sp}$ at a fixed frequency of 6 GHz. The fitting results reveal that the voltage has a $\sin^2\varphi$ dependence on the precession cone angle.

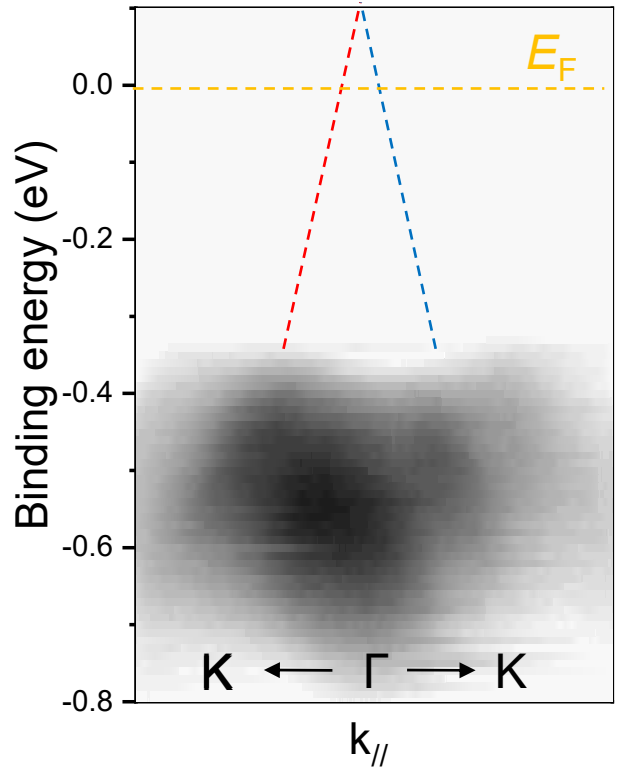

**Figure S7.** ARPES result of the BST films at room temperature. The ARPES measurement was performed at room temperature along the K-Γ-K direction with a SPECS PHOIBOS 150 hemisphere analyzer. The bulk valance band with a bulk energy gap is clearly seen in the spectrum, in agreement with the previous report.[5] The red and blue dashed lines indicate the Dirac surface states with opposite spin polarities, which intersect at the Dirac point above the Fermi level ($E_F$, represented by the yellow dashed line). The intrinsic defects act as hole doping at room temperature, resulting in the topological surface states obscure. Additionally, transferring the sample from the PLD chamber into the ARPES system may induce surface contamination.

Table S1. Summary of the $\lambda_{IEE}$ of TIs determined by SP-FMR.[6-15]

| Heterostructure | Growth Method | $\lambda_{IEE}$ (nm) | Ref. |
| --- | --- | --- | --- |
| $Bi_xSe_{1-x}$/YIG | Sputtering | 0.11 | [6] |
| $Bi_2Se_3$/YIG | Sputtering | $2.2 \times 10^{-3}$ | [7] |
| $(Bi,Sb)_2Te_3$/YIG | MBE | $35 \times 10^{-3}$ | [8] |
| $Cr-(Bi,Sb)_2Te_3$/YIG | MBE | $37 \times 10^{-3}$ | [8] |
| $BiSbTe_{1.5}Se_{1.5}$/Cu/NiFe | PLD | 0.8 | [9] |
| $Bi_2Se_3$/Bi/Fe | MBE | 0.28 | [10] |
| $Bi_xSe_{1-x}$/CoFeB | Sputtering | 0.32 | [11] |
| $Sb_2Te_3$/Au/Co/Au | MOCVD | 0.61 | [12] |
| $\alpha$-Sn/Ag/Fe | MBE | 2.1 | [13] |
| CdTe/HgTe/HgCdTe/NiFe | MBE | 2.0 | [14] |
| $Sb_2Te_3/Bi_2Te_3$/Au/Co/Au | MOCVD | 0.44 | [15] |
| $(Bi,Sb)_2Te_3$/YIG | PLD | 4.75−5.6 | This work |

MBE: molecular beam epitaxy; MOCVD: metal-organic chemical vapor deposition